\documentclass[12pt,epsf]{article}
\usepackage{graphicx}

\begin{document}

\sloppy
\begin{flushright}{SIT-HEP/TM-24}
\end{flushright}
\vskip 1.5 truecm
\centerline{\large{\bf Brane necklaces and brane coils }}

\vskip .75 truecm
\centerline{\bf Tomohiro Matsuda
\footnote{matsuda@sit.ac.jp}}
\vskip .4 truecm
\centerline {\it Laboratory of Physics, Saitama Institute of
 Technology,}
\centerline {\it Fusaiji, Okabe-machi, Saitama 369-0293, 
Japan}
\vskip 1. truecm
\makeatletter
\@addtoreset{equation}{section}
\def\theequation{\thesection.\arabic{equation}}
\makeatother
\vskip 1. truecm

\begin{abstract}
\hspace*{\parindent}
We investigate the evolution of the networks of the cosmic strings in
 angled brane inflation. 
We show how they can be distinguished from the conventional ones.
The cosmic strings in angled inflation are
the daughter $D_{p-2}$ branes that are extended between the mother
$D_p$ branes.
In the effective action, the strings should have a moduli,
since the endpoints of the $D_{p-2}$ branes can move
freely on the $D_p$ branes. 
Then naturally the position of the $D_{p-2}$ branes, which corresponds to
the moduli of the $(1+1)$-dimensional effective action,
can vary along the cosmic strings.
The variation of the moduli results in 
the peculiar $(1+1)$-dimensional kink configurations.
The kinks are the monopoles on the strings.
Therefore, the cosmic strings in angled inflation become necklaces.
The loops of the necklaces can shrink to produce stable
 winding states, which look like coils.
We show why the cosmological implications of the brane necklaces 
are  important.
We point out that the cosmic strings in generic models of brane inflation
should become necklaces, depending on the structure of the compactified
 space and the effective potential  of the model. 
\end{abstract}

\newpage
\section{Introduction}
Models with more than four dimensions are interesting, because all the
physical ingredients of the Universe can be unified in a higher
dimensional theory.
String theory is the most promising scenario where quantum
gravity is included by the requirement of additional dimensions and
supersymmetry.
The idea of large extra dimension\cite{Extra_1} is important,
because it may solve the hierarchy problem.
In this case, the observed Planck mass is obtained by the relation
$M_p^2=M^{n+2}_{*}V_n$, where $M_{*}$ and $V_n$ denote the fundamental
scale of gravity and the volume of the $n$-dimensional compact space,
respectively. 
In the scenarios of large extra dimension, the fields in the standard model
are expected to be localized on wall-like structures (``our branes''), while
the graviton propagates in the bulk. 
In the context of string theory, a natural embedding of
this picture is realized by brane construction.
The brane models are interesting from both phenomenological and 
cosmological viewpoints.

Analyses of cosmological defect formation are important in brane models.
\footnote{Inflation in models of low fundamental scale 
are interesting\cite{low_inflation, matsuda_defectinfla, matsuda_nontach}.
Scenarios of baryogenesis in such models are discussed in ref.
\cite{low_baryo, Defect-baryo-largeextra, 
Defect-baryo-4D}, where defects play important roles.
Brane defects such as monopoles, strings, domain walls and
Q-balls\cite{BraneQball, matsuda_monopoles_and_walls} are important
because we are expecting that future cosmological observation will reveal
the cosmological evolution of the Universe, which will also reveal
the physics beyond the standard model.
If one wants to know what kinds of brane defect are produced
in the early Universe, one needs to understand how they are formed.}
For the cosmological scenarios, one can consider at least three
different types of defects:
\begin{itemize}
\item Defects are branes.\\
      In this case, cosmological defects are the branes that
      have less than three spatial dimensions in the uncompactified
      space-time.
      In the previous discussions\cite{angled-defect}, in which the
      dynamical effect of ``our brane'' has been neglected, it was
      concluded that the cosmological production of
      monopoles and domain walls are negligible in the models of
      brane inflation. 
      Later in ref.\cite{matsuda_monopoles_and_walls} and
      \cite{simulation_braneproduction},  however, it was
      shown that cosmological 
      formation of such monopoles and domain walls are quite natural in
      generic cosmological scenarios.
\item Defects are deformed branes.\\
      In this case, the cosmological defects are formed by the continuous
      deformation of branes.
      First in ref.\cite{Alice-string}, and later in
      ref.\cite{matsuda_monopoles_and_walls,incidental}, it was shown that
      the fields that parameterize the position of branes can
      fluctuate in spatial directions to form cosmological brane defects. 
      Cosmic strings are constructed in ref.\cite{Alice-string}, where
      the singularity in the core is resolved by smearing the wall-like
      structure.
      Monopoles, strings and domain walls are constructed in
      \cite{matsuda_monopoles_and_walls, incidental}. \footnote{It is
      possible to construct branes as defects 
      in higher dimensional gauge
      theories. Interesting
      applications of this idea are discussed in ref.\cite{TIT-group}.} 
\item Localized fields are shifted in the cosmological defects.\\ 
      In ref.\cite{Localize_on_fat_wall}, localization of matter
      fields on a fat domain wall is discussed to explain small
      interactions in the effective four-dimensional action.
      This idea is important in constructing realistic models in which
      proton must be stabilized.
      Along the line of this argument,
      new defect configurations are constructed in
      ref.\cite{Defect-baryo-largeextra}, which induce shifts of the
      localization.
      Due to the shifts of the localization, small interactions can be
      enhanced in the core of these defects.
      Therefore, these defects can assist the generation of baryon
      number asymmetry of the Universe\cite{Defect-baryo-largeextra,
      Defect-baryo-4D}. 
\end{itemize}

First, we shall review the essence of the ``previous
arguments''\cite{angled-defect}, 
which were used to exclude the cosmological production of monopoles and
domain walls.
In the original scenario of brane inflation\cite{brane-inflation0},
inflationary expansion is driven by the potential between D-brane
and anti D-brane evolving in the bulk.
Then the scenario of the inflating branes at a fixed angle is studied
in ref.\cite{angled-inflation}, where the slow-roll condition is 
improved by introducing a small angle.
The end of brane inflation is induced by the brane
collision where brane annihilation (or recombination) proceeds through
tachyon condensation\cite{tachyon0}.  
During brane inflation, tachyon is trapped in the false vacuum, which
can result in the formation of lower-dimensional branes after brane
inflation. 
The production of cosmological brane defects was discussed
in ref.\cite{angled-defect}, where it was concluded
that cosmic strings 
are copiously produced in these scenarios, while monopoles and domain
walls are negligible. 
To be precise, their arguments are based on the speculation that
any variation of fields in the compactified direction must be suppressed
during inflation since the compactification radius must be small 
compared to the horizon size during inflation.
Then they have concluded that the daughter branes, which are formed by
tachyon condensation on the world volume of the mother branes, must wrap
the same compactified space as the mother branes. 
In this case the effect of compactification seems significant.
As a result, it was concluded that the codimension of the daughter branes
must lie within the uncompactified space.
Since the number of the codimension must be even in this case, the
cosmological defects must be cosmic strings.
Moreover, later in ref.\cite{RR-dvali-sting}, it was discussed that the
analysis 
does not fully account for the effects of compactification, because the
directions transverse to the mother branes had not been considered.
The effect of the RR fields extended
to the compactified dimensions was discussed in ref.\cite{RR-dvali-sting}.
Then it was concluded that the creation of the gradients of the RR fields
in the bulk of the compactified space is
costly in energy, so that there should be a serious suppression in the 
creation of the daughter D brane as far as they do not fill all the
compactified dimensions. 
As a result, in the ``previous arguments'', cosmological brane defects
were discussed to suffer from many unnatural constraints.

However, here we should recall that no such suppression has been
discussed so far in the cosmological production of the conventional defects
in four-dimensional effective gauge theory, while the realistic brane
models must reproduce the standard model in its effective action.  
Therefore, it is quite natural to think about the following questions.
Is it really impossible to produce branes that do not
wrap the same compactified space as the mother branes?
Is it really impossible to produce monopoles and domain walls in brane
inflation? 
Here we should note that the ``previous arguments'' are not fully reliable
in more realistic cases.
The most obvious example was first discussed in
ref.\cite{matsuda_angleddefect}\footnote{See also
ref. \cite{matsuda_monopoles_and_walls}},  
where the inconsistency of the string tension is solved 
in angled inflation.
In the original argument in ref.\cite{angled-defect}, the
tension of the string did not coincide with the one calculated from the
effective action.
This inconsistency is clearly due to the assumption that the daughter
brane must 
 wrap the same compactified space as the mother brane. 
In fact the previous arguments might be true in some simplest cases,
however we must be more careful about the process of the brane
recombination.
In ref.\cite{matsuda_monopoles_and_walls,matsuda_angleddefect},
it was shown that the production of the cosmic strings in angled inflation
is realized by the creation of the daughter branes 
that are {\bf extended} between the splitting mother branes. 
It should be noted that the creation of such extended branes is not a
counter example to the generic mechanism of tachyon condensation.
To be precise, one can understand from a careful treatment of the
effective action\cite{Localize_tachyon} that the eigenfunction of the
tachyonic mode should be localized at the intersection.
Since the mechanism of this localization is {\bf different} from the
Kibble mechanism, this argument does not contradict to the ``previous
arguments''.
Therefore, the ``seed'' must be localized at the
intersection.   
As the recombination proceeds, the $D_{p-2}$ brane at the intersection is
pulled out of the mother $D_p$ branes, and finally becomes extended
between mother branes.
In this case the serious problem of the RR field is also avoided, since the
length of the extended daughter brane vanishes when it is created at the
intersection.
It costs energy to pull $D_{p-2}$ branes out of the $D_p$ branes,
however the cost is paid by the repulsive force between the
splitting $D_p$ branes.
Of course the tension of such extended daughter branes
matches precisely to the D-term strings in the effective action.
Moreover, this conclusion is consistent with the analysis of the string
production in the 
effective action, where the production of the D-term strings is not
suppressed. 
Of course, it seems obvious from the string perspective that the
brane creation is due to the local dynamics of the open strings at the
intersection of the splitting $D_p$ branes, which suggests the obvious
localization of the eigenfunction.
The daughter branes satisfies the required properties.
Thus at least in the models of angled inflation, it is obvious that the
cosmic strings in the effective action do not wrap the same compactified
space as the mother branes. 

From the above discussions, it is natural to
think that the previous arguments are not reliable and should be
compensated in more generic cosmological scenarios.
Obviously, there is no reason that we must believe that in their final
state the daughter branes must wrap the same compactified space as the mother
branes.
Since the previous argument cannot fully account for the generic
process of the cosmological defect formation, one can hardly accept their
conclusions.
Is it really impossible to produce sufficient amount of cosmic
monopoles and domain walls by the brane creation?
This question is answered in ref.\cite{matsuda_monopoles_and_walls,
matsuda_angleddefect}, paying careful attention to the brane dynamics
after inflation, especially to the dynamics of ``our brane''. 
For the extended branes to be produced between mother branes, 
the distance between mother branes is required to be zero at a
moment.
This idea is generic and applicable to other conventional
cosmological processes. 
The spontaneous symmetry breaking in the effective action is sometimes
described by the recombination of the 
branes\cite{Intersecting_Brane_Models} or by the branes falling apart,
which can be induced by the thermal
effects\cite{thermal-brane}.
As a result, contrary to the previous speculation, it was
shown\cite{matsuda_monopoles_and_walls} that there is no reason that one
should believe that monopoles and domain
walls\cite{Matsuda-weak-wall} are suppressed in the brane Universe, once
the dynamics of ``our brane'' is included. 

Besides the cosmological defects that are formed by the  brane creation,
one can consider the defects that are formed
by the continuous deformation of the branes.
The two kinds of the brane defects can be produced 
by the same process\cite{matsuda_monopoles_and_walls, incidental}. 
Therefore, the actual cosmological defects should be {\bf the mixture of
these defects.}
It should be noted that the analyses of such mixed defects 
are quite important in understanding the evolution of the brane Universe.

\begin{figure}[ht]
 \begin{center}
\begin{picture}(410,340)(0,0)
\resizebox{13cm}{!}{\includegraphics{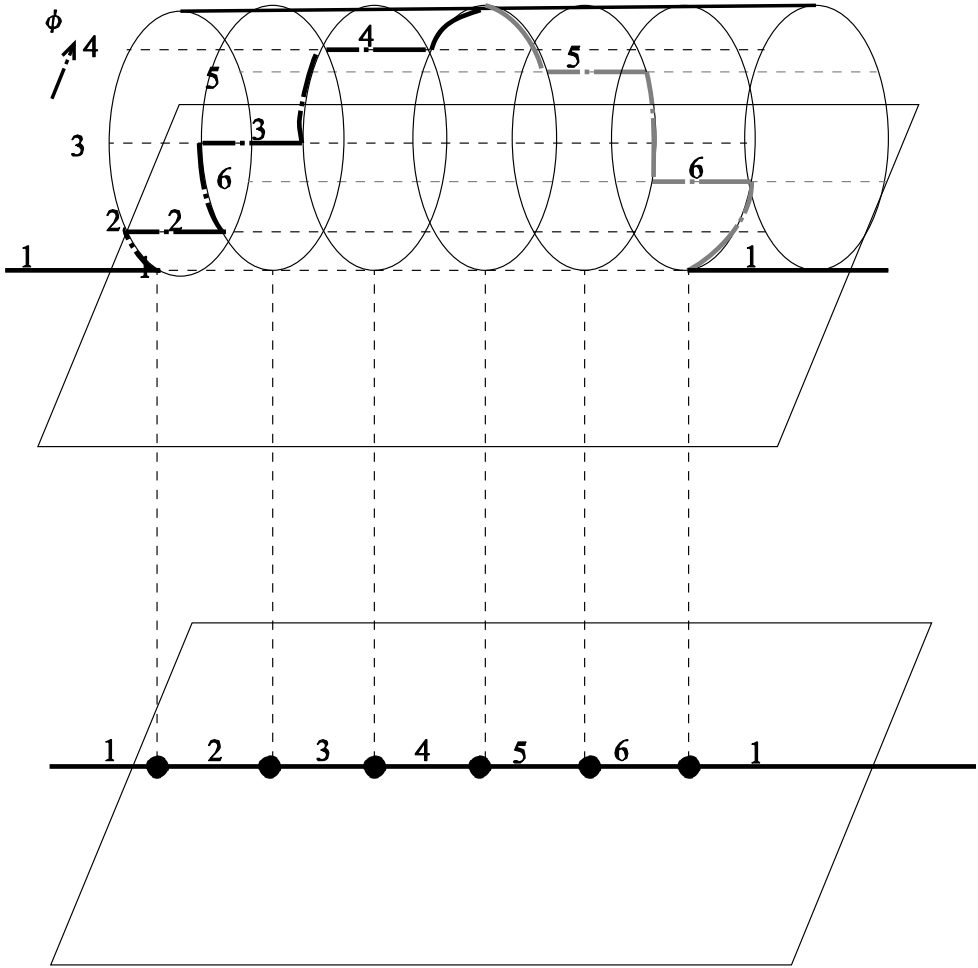}} 
\end{picture}
\caption{Strings in angled inflation could
  have many domains. The domains are interpolated by the kinks on the
  strings. Since the kinks are the monopoles skewered with the strings, the
  strings become brane necklaces.} 
\label{fig:necklace}
 \end{center}
\end{figure}

\begin{figure}[ht]
 \begin{center}
\begin{picture}(410,300)(0,0)
\resizebox{15cm}{!}{\includegraphics{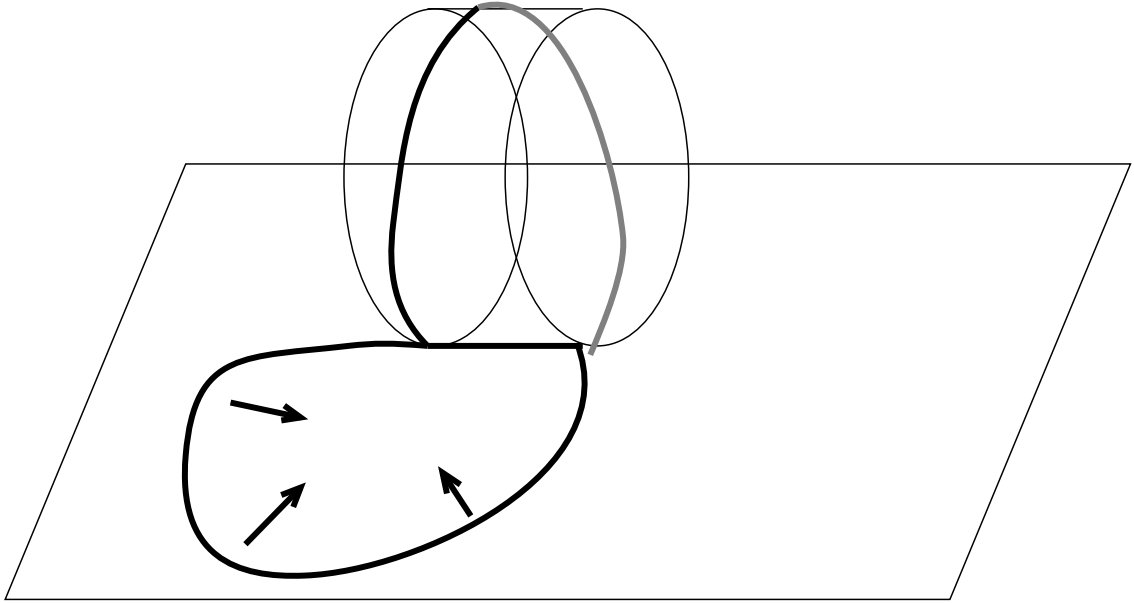}} 
\end{picture}
\caption{A chopped loop of the brane necklace can shrink. However,
  it cannot shrink to a point if the loop winds around compactified
  space.} 
\label{fig:shrink}
 \end{center}
\end{figure}

From the above viewpoints, we shall reconsider the evolution
of the cosmic strings in angled inflation.
In this paper, we show that the strings in angled inflation can be
distinguished from the conventional cosmic strings.
In angled inflation, cosmic strings are the
daughter $D_{p-2}$ branes that are extended between mother $D_p$ branes.
The $D_{p-2}$ branes have the flat direction (moduli) in the compactified
space, along which the endpoints of the $D_{p-2}$ branes can move
 freely on the $D_p$ branes.\footnote{See fig.\ref{fig:flat}}
Then the position of the $D_{p-2}$ branes can vary along the
 cosmic strings, which results in the peculiar type of
 $(1+1)$-dimensional  kink configurations.
One can also find the peculiar winding states that are defined on the world
volume of the strings. 
A schematic picture of the kinks is shown in
fig.\ref{fig:necklace}.
The point-like objects that appear on strings are monopoles.
Since the monopoles are skewered with the strings, the 
cosmic strings in angled inflation 
become necklaces.
These kinks are produced by the spatial
deformation of the $D_{p-2}$ branes.
Therefore, the brane necklaces are the {\bf hybrid} of
the brane creation and the brane deformation.
The chopped loops of the necklaces can shrink to produce the stable winding 
states, which looks like coils.
A schematic picture of the brane coils is shown in fig.\ref{fig:shrink}.
It should be noted here that the brane coils wind around the
 compactified space that is {\bf different} from the mother branes.
Of course the mechanism that induces such windings is different from 
the Kibble mechanism.
Therefore, our mechanism avoids the serious constraints that have been discussed previously
in ref.\cite{angled-defect}.

\section{Brane necklaces}
In the early Universe there could be a pair of brane anti-brane
separated in the compactified space.
The potential energy of these branes can drive brane inflation,
then they annihilate rapidly at the end.
If this process contains only a pair of $D_p\overline{D}_p$
branes, it might be true that only cosmic strings are
produced after inflation.\footnote{Recently, it was shown by numerical
simulations that many kinds of branes 
may appear in the intermediate state, which may annihilate to produce
stable monopoles and domain walls in their final
state\cite{simulation_braneproduction}. } 
In this case, the cosmic strings are the daughter branes that wrap
the same compactified space as the mother branes.
At first the strings could move freely in the compactified space,
but later their position should be fixed as was discussed in
ref.\cite{Polchinski-GW, vilenkin-lowp}. 
Then, if one assumes that other branes (including ``our branes'') are 
decoupled, these strings cannot break nor
dissolve 
into other branes.
In this case, since other branes are completely decoupled from
the cosmic strings, the interactions of the cosmic strings
must be quite simplified.
The cosmological consequence of this scenario is already investigated in
ref.\cite{Polchinski-GW, vilenkin-lowp}.

On the other hand, in generic cases it seems rather difficult to obtain
enough number of ``decoupled'' cosmic strings while reheating 
the Universe to an acceptable temperature.
If the moving brane of the brane inflation collides to the stack of
the Standard Model (SM) branes, reheating should be successful.
However, in this case one cannot ignore the interaction between
the daughter branes and the other branes at the stack.
Since the cosmic strings are the lower-dimensional branes that
could break or dissolve into other branes,  in generic cases one cannot
simply ignore the interaction between other branes.
Therefore, it is important to investigate cosmic strings that
intersect with other branes.

\begin{figure}[ht]
 \begin{center}
\begin{picture}(410,300)(0,0)
\resizebox{15cm}{!}{\includegraphics{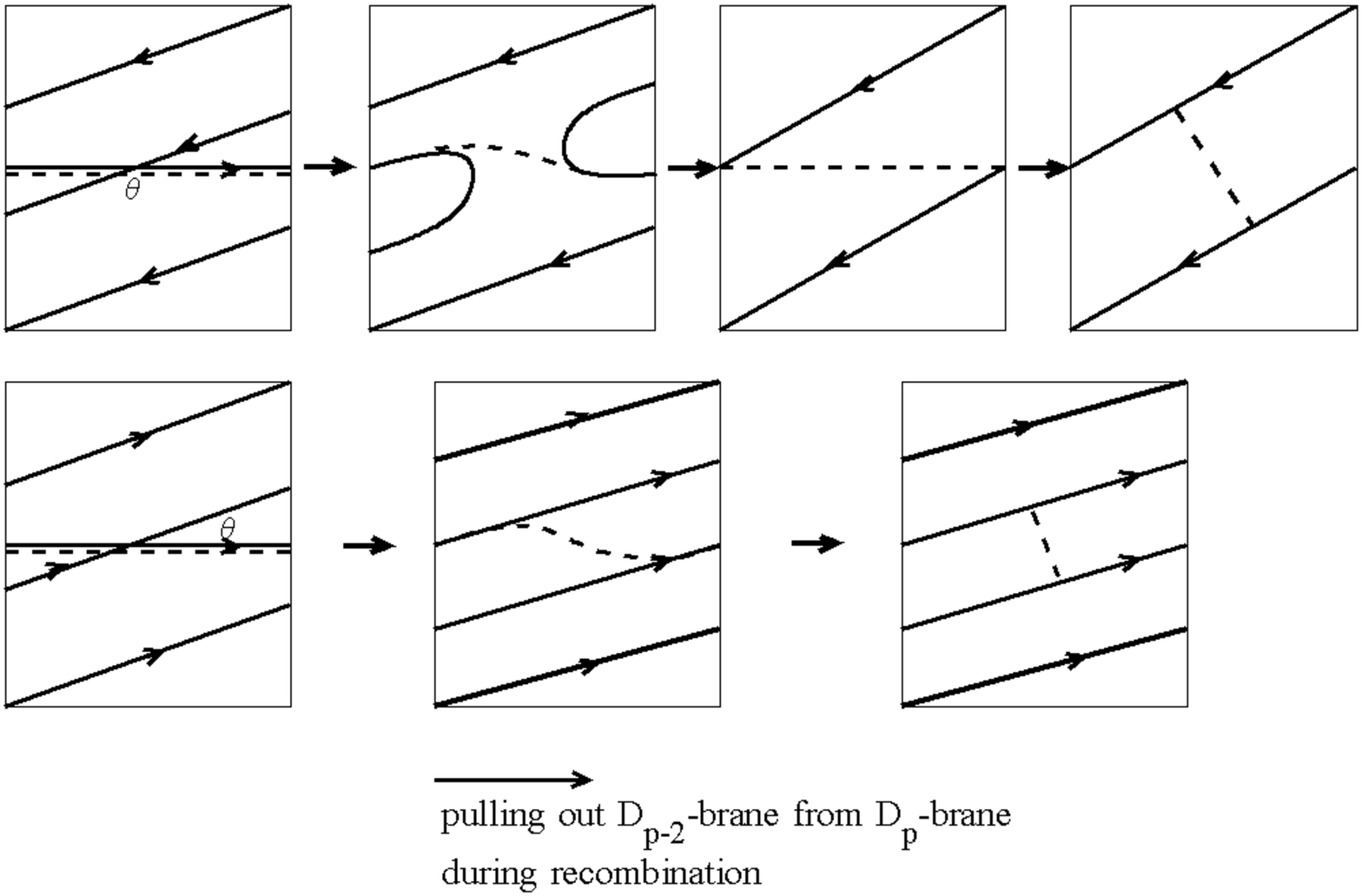}} 
\end{picture}
\caption{Upper row: schematic recombination of two $D_p$-branes with
  $(\pi-\theta) \ll 1$. The dashed line on the
  $D_p$-brane represents the $D_{p-2}$-brane that might appear on the
  worldvolume of the $D_p$-brane when the tachyon condenses.
As the recombination proceeds, the $D_{p-2}$ brane is
pulled out from the mother brane, and finally becomes extended
between the mother brane.
  Second row: schematic recombination of two $D_p$-branes with 
  $\theta \ll 1$. 
In both cases the daughter $D_{p-2}$ brane does not wrap the same
  compactified space as the mother brane.}
\label{fig:split}
 \end{center}
\end{figure}

From the above viewpoints, we shall try to understand the evolution of
the cosmic strings in angled inflation to show how one can distinguish
the cosmic strings in angled inflation from the conventional ones.
As we have discussed in the previous section, angled inflation is the
most obvious generalization of the simplest 
brane inflation, where the slow-roll condition is improved
and the {\bf cosmic strings intersect with ``our branes''}. 
If the potential that lifts the flat direction of the
cosmic strings is shallow, the Brownian motion in the compactified space 
must be important. 
The Brownian motion in the direction
that is perpendicular to the daughter brane can be
induced by the brane collision\cite{braun_string}.
Even if the potential is not shallow due to the stringy
effects, 
it seems natural to expect that the internal position of the cosmic strings
can vary because of the energetic process of the brane
collision. 
In any case, it seems too optimistic to expect that the potential is
so steep that one can avoid the generation of the kinks on the
strings. 
Moreover, if the angle is small($\theta\ll 1$) and the final state of
the recombined mother brane wraps about
$\theta^{-1}$ times around the compactified space ,\footnote{See
fig.\ref{fig:split}} the potential should have many local minima.
These minima appear as the ``domains'' on the strings.
To be precise, strings can have several distinctive
regions(domains) where the branes are located at different sites in the
compactified 
space.\footnote{See fig.\ref{fig:necklace}} 
Then kinks (i.e. (1+1)-dimensional domain walls) will appear on the
strings, which interpolates between neighboring domains on the strings.
These kinks may have interesting consequences.\footnote{Here we should
note that the ``kinks'' on the strings are 
different from the ``kinks and cusps'' that are sometimes
used in the discussion about the decay of the conventional cosmic strings.}
Moreover, it is interesting to think about the configuration where the
internal position of the cosmic string is changed so that it  
finally wraps around a non-trivial circle of the compactified space.
Then a peculiar type of topological number can be defined {\bf on} the
strings, which can stabilize the chopped loops.\footnote{See
fig.\ref{fig:shrink}.}  
The winding states of the strings look like coils.
Even though the brane coils are stabilized by their windings, they can
be annihilated with another coils that have the opposite windings\footnote{As
is already discussed in 
ref.\cite{Alice-string,incidental},
branes might be delocalized (smeared) in the core of a defect.
If such delocalization is energetically favored, the loops around
compactified space could be unwinded.
In our model, however, it is obvious that the process is corresponding
to the instanton-like phase transition, which is suppressed by the
significant exponential factor.}

The cosmological evolution of the conventional necklaces is already
investigated 
in ref.\cite{vilenkin-necklace}.\footnote{BPS necklaces in SQCD
are discussed in ref.\cite{TIT-group, tong}.}
Therefore, it is quite interesting to investigate the brane
necklaces so that we can find the crucial difference from the
conventional necklaces. 
Because of their peculiarity, the brane necklaces can be
used to distinguish the brane Universe from the conventional one.
Of course the above speculation seems rather naive, therefore
we shall discuss this issue in more detail.

\subsection{String production after angled brane inflation}
\begin{figure}[ht]
 \begin{center}
\begin{picture}(310,190)(0,0)
\resizebox{8cm}{!}{\includegraphics{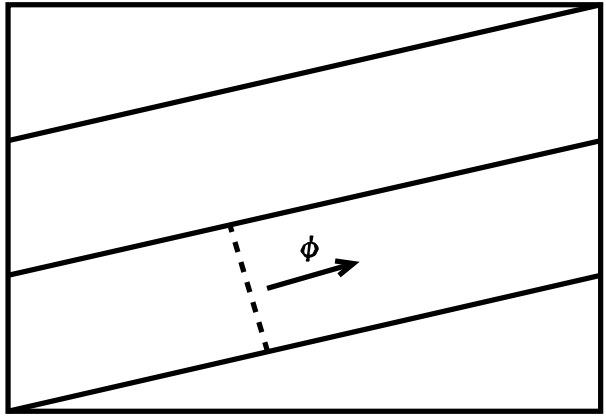}} 
\end{picture}
\caption{There is a direction in the compactified space along which the
  branes(=strings) can move. 
 The flat direction is denoted by $\phi$.}
\label{fig:flat}
 \end{center}
\end{figure}

To make our discussions simple and convincing, here we consider angled
brane inflation with a small angle($\theta \ll 1$).
In this case, the eigenfunction of the tachyonic mode should be localized at 
the intersection.
It should be noted that this mechanism of the localization is {\bf
different} from the conventional Kibble mechanism, which suggests that our
result does not contradict to the previous arguments.
Moreover, comparing the effective tension of the daughter branes with
the one obtained in the effective action, one can easily confirm that the 
cosmic strings in angled inflation must be extended between mother 
branes.\footnote{See fig.\ref{fig:split}.} 
If the daughter branes are extended between the mother branes as is shown
in fig.\ref{fig:flat}, it is easy to see
that the daughter branes can move freely along the mother branes.
The flat direction that corresponds to the free motion in the compactified
space is depicted in fig.\ref{fig:flat}.
Of course, we know that the flat direction is supposed to be
lifted by the potential that is due to the conventional supersymmetry
breaking, which induces the mass of $O(m_{3/2})$. 
On the other hand, since we are considering small angle $\theta \ll 1$,
the flat direction $\phi$ is periodic and winds about $\sim \theta^{-1}$
times around the compactified space.
Therefore, the number of (local or global) minima should be as large as 
$\theta^{-1} \gg 1$.  
If the potential that lifts the flat direction is shallow, the brane
strings can initially move freely along the flat direction.
Even if the potential is not so shallow due to the stringy effects,
one can naturally expect that the collision is so
energetic that the daughter branes can stay at any minima.
Branes located at different sites in the compactified space cannot
intersect with each other\cite{Polchinski-GW, vilenkin-lowp}, because their
separation in the compactified space is larger than the string scale.  

\begin{figure}[ht]
 \begin{center}
\begin{picture}(410,200)(0,0)
\resizebox{10cm}{!}{\includegraphics{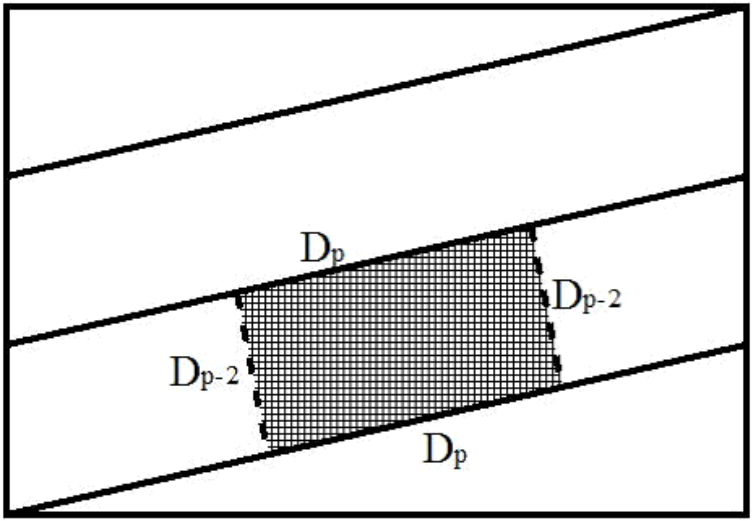}} 
\end{picture}
\caption{A $D_{p-2}$ brane is expanded in the (p-2)-dimensional
  compactified space. The 
  boundary of the $D_{p-2}$ brane is $D_p$ and $D_{p-2}$ branes.}
\label{fig:monopole}
 \end{center}
\end{figure}

\begin{figure}[ht]
 \begin{center}
\begin{picture}(400,280)(0,0)
\resizebox{10cm}{!}{\includegraphics{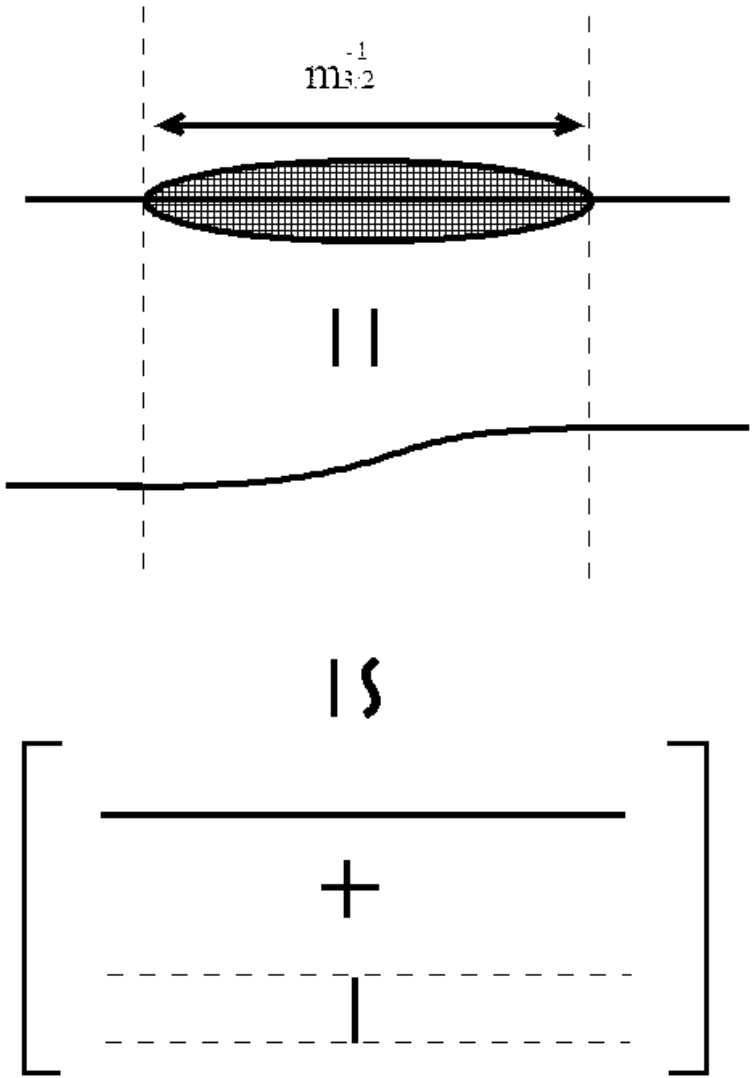}} 
\end{picture}
\caption{A fat monopole is skewered with a string. }
\label{fig:skew}
 \end{center}
\end{figure}

Let us consider the $(1+1)$ dimensional world on the strings.
The $(1+1)$ dimensional world on the strings will have many domains of 
the number of $\sim\theta^{-1}$, which correspond to the degenerated or
quasi-degenerated vacua on the flat direction.
Then the domain walls will appear on the strings, which interpolate
between the two neighboring domains.\footnote{See 
fig.\ref{fig:necklace}.} 
The width of the domain walls is determined by the shape of the
effective potential $V(\phi)$, 
which becomes $\delta_m \sim m_{3/2}^{-1}$ if the potential is induced by the
conventional mechanism 
of supersymmetry breaking, or can become as thin as $\delta_m \sim M_*^{-1}$
if the potential is lifted by the stringy dynamics.
\begin{itemize}
\item
In the former case, the height of the effective potential is $V_0 \simeq
m_{3/2} \Delta \phi$, where $\Delta \phi 
\simeq R_E M_*^2$ denotes the typical distance between the neighboring
vacua. 
From the dimensional argument one can easily understand that the mass of
     the monopoles is $m \sim \Delta \phi$.\footnote{It 
should be noted that the effective  
action of the $(1+1)$ dimensional world of the strings must have
non-zero cosmological constant due to the tension of the
string in the background.
Therefore, the precise formula of the effective potential should contain
the effective cosmological constant, which may or may not dominate the
     vacuum energy.
Since the cosmological constant in this case is nothing but the tension
of the strings, one can easily understand that kinks are point-like
objects that are ``skewered'' with the strings.
On the other hand, if one considers point-like objects to which two
strings are 
``attached'', the definition of the mass should be obscured because of
     the contribution from the string tension.
In this paper, we shall consider the former (skewered) picture for 
     fat monopoles. See also fig.\ref{fig:skew}}
\item
From the brane perspective the kinks are corresponding to the $D_{p-2}$
branes that are extended along the internal 
direction.\footnote{See fig.\ref{fig:monopole}.}
The kinks are nothing but the monopoles, which correspond to the
$D_{p-2}$ branes that are extended in the $(p-2)$-dimensional compactified
space.
Then one can easily calculate the effective mass of this brane
configuration, which becomes $m\sim M_*^3 R_E^2 \theta$.
\end{itemize}

As we have discussed above, there are two different contributions 
to the mass of the monopoles.
When one considers the SQCD-MQCD correspondence, one should think that 
the potential in the effective action must properly represent the brane
dynamics. 
Of course, in such cases the two must agree\cite{tong}.
On the other hand, if the effective potential that lifts the
flat direction is due to the undefined mechanism of supersymmetry
breaking, the results cannot agree.
Therefore, in our case the mass of the monopoles becomes $m\sim \Delta
\phi$ in the limit $\theta \rightarrow 0$, while
the mass is dominated by the brane tension if $1\gg \theta > (M_*R_E)^{-1}$.
In this paper we shall consider these two different limits without
specifying the origin of the effective potential.

\subsection{Stability and decay of brane necklaces}

Let us consider the evolution of the network of the brane necklaces.
The strings at different domains cannot reconnect because they are
placed at a distance in the compactified space.
Therefore, in the length scale that is much larger than the distance between
the monopoles, the reconnection probability of the necklaces is suppressed by
the number of the minima:
\begin{equation}
p\simeq \theta
\end{equation}
The evolution of such strings is already discussed in
ref.\cite{vilenkin-lowp}. 
According to ref.\cite{vilenkin-lowp}, one can easily understand that
the cosmic strings in $\theta \ll 1$ angled inflation are quite important in
the observation of the gravitational wave signals.
However, here we must be cautious because the evolution of the network
of the necklaces can be different from the conventional  $p\ll1$ strings.
Therefore, we should first examine the effect of the monopoles that might or
might not affect the evolution of the network of the $p\ll 1$
strings.
Of course, it is apparent that the details must be investigated using 
numerical simulations, however here we can modestly assume that one can
use the conventional analyses as far as the parameter space of the model
is within the range of ref.\cite{vilenkin-lowp, vilenkin-necklace}.

To make our arguments clear, we use the same notations as in
ref.\cite{vilenkin-lowp, vilenkin-necklace}.
In ref.\cite{vilenkin-necklace}, it was assumed that the monopoles are
formed {\bf before} the strings are produced at the second phase transition.
In our case, however, the monopoles are formed {\bf after} the strings are
formed.
Since the succeeding evolution of the necklaces is
irrelevant to how they are formed, the analysis in
ref.\cite{vilenkin-necklace} is applicable in our case.
The sequence of the phase transition is only relevant to
their initial condition.
For example, if monopoles are produced long after strings are produced,
their initial average separation $d$ will be large compared to the original
analysis of the conventional necklaces.
\begin{itemize}
\item
If the potential is lifted by the conventional mechanism of
supersymmetry breaking, the temperature $T_m$ when monopoles are formed
can be estimated as $T_m \sim (m_{3/2}M_p)^{1/2}$.
In this case, the width of the monopoles $\delta_m$ is as large as
$\delta_m \sim m_{3/2}^{-1}$.
Therefore, their initial average separation $d$ is inevitability as
large as the horizon size.
\item
As we have stated above, one can consider steep potential that could be
induced by the stringy effects.
In this case, monopoles should be produced at the same time when strings are
produced, which suggests that the initial average separation between 
monopoles can range from the minimal distance
$d\sim M_*^{-1}$\cite{simulation_braneproduction} 
to the horizon size. 
\end{itemize}

Here we introduce an important quantity for the evolution of the necklaces,
which is the dimensionless ratio $r=m/(\mu d)$.
Here $\mu$ and $d$ denote the tension of the strings and the separation
between the monopoles, respectively.
In general, the initial value of $r$ can be large or small, depending on
the nature of the two phase transitions.
As we have discussed above, we are considering two different limits.
\begin{itemize}
\item When monopoles are thin and are dominated by the brane dynamics,
the mass of the monopoles is $m \simeq M_*^3 R_E^2 \theta$.
In this case we can assume $d > M_*^{-1}$,
which results in the initial value of the ratio $r_{ini} < M_* R_E$.
\item On the other hand, if the potential is almost flat and is lifted by the
conventional mechanism of supersymmetry breaking, the mass of the
monopoles becomes $m \simeq \Delta \phi$.
In this case, since the width of the monopoles is as large as
$m_{3/2}^{-1}$, we must assume $d > m_{3/2}^{-1}$.
Therefore, the initial value of the ratio is $r_{ini} < m_{3/2} /(M_* \theta)$.
\end{itemize}
In both cases we can put a modest assumption that $r_{ini} <1$.
Following the arguments in ref.\cite{vilenkin-necklace},
we can obtain the following equation for $r(t)$:
\begin{equation}
\frac{\dot{r}}{r}=-\frac{\kappa_s}{t}+\frac{\kappa_g}{t},
\end{equation}
where the first term on the right hand side describes the string
stretching due to expansion of the Universe, while the second term
describes the competing effect of string shrinking due to gravitational
radiation.
In this regime, one can find that if $r$ is
initially small it will grow at least until $r\sim 1$.
As $r$ grows, monopole anti-monopole (or coil anti-coil) annihilation
should become important and the growth of $r$ will terminate.
\begin{itemize}
\item If the potential is steep and the mass of the monopoles is
given by $m \simeq M_*^3 R_E^2\theta$, the width of the
strings $\delta_s$ should be comparable to the width of the monopoles
$\delta_m$.
Therefore, in this case one can assume that $\delta_s \sim \delta_m\sim
M_*^{-1}$, which leads to the result $r_{max}\simeq R_E M_*$.
This result suggests that as far as $ R_E M_* <10^{6}$ the
parameter space of our model is within the 
range of the original argument in ref.\cite{vilenkin-necklace}.
If $r\gg 1$, the loop self-intersections should be frequent and their
fragmentation into smaller loops is very efficient.
Therefore, a loop of size $l$ typically disintegrates on a time scale
\begin{equation}
\tau_{r\gg1} \sim \frac{l}{\sqrt{r}},
\end{equation}
which modifies the conventional analysis of the $p\ll 1$ strings.
\item If the potential is almost flat and the mass of the monopoles is given
by $m \simeq \Delta \phi$, the maximum value of the ratio becomes
$r_{max} \simeq m_{3/2}/(M_* \theta)^{-1} \ll 1$, which suggests that
the monopoles are negligible.
In this case, the standard evolution of the string network is applicable. 
\end{itemize}

Let us recall the main quantities that are used in the standard analysis
of the evolution of a string network.
The long string network is characterized by the parameters
$\xi(t)$, $L(t)$, and $l_{wiggles}(t)$.
Here $\xi(t)$ is the coherence length, which is defined as the distance
beyond which the directions along the string are uncorrelated.
$L(t)$ is the average distance between the strings, and $l_{wiggles}(t)$
is the characteristic wavelength of the smallest wiggles that appear on
long strings.
The length $l(t)$ of the chopped-off loops are characterized by the
parameter $\alpha$,
\begin{equation}
l(t)=\alpha t,
\end{equation}
where the standard value of $\alpha$ is determined by the gravitational
radiation losses from loops, which is given by 
\begin{equation}
\alpha^{st}\sim \Gamma G \mu,
\end{equation}
where $\Gamma$ is a numeriacl coefficiend of O(10).
If the gravitational radiation from counter-streaming
wiggles on long strings is much less efficient in damping the wiggles
than ordinarily thought, $\alpha$ should be much smaller than
$\alpha^{st}$\cite{lessalpha}. 
In our case, if the ratio $r$ is small, the
equation of motion is not different from the original argument,
which means that 
we can examine the evolution of the network just using the results
obtained in
ref.\cite{vilenkin-lowp}.
In the macroscopic scale where the reconnection probability is small,
one can estimate the above parameters as
$L(t)\sim p^{1/2} t$ and $ \xi(t)\sim t$.
A significant fraction of the total string length within Hubble volume
should go into loops each Hubble time, whose number density is
\begin{equation}
n(t) \sim \frac{1}{p\Gamma G \mu t^3},
\end{equation}
which is precisely the same as the original result.

However, what we should show in this paper is the difference from the
standard results, which can be
used to distinguish the brane necklaces from the conventional defects. 
In the original argument in ref.\cite{vilenkin-lowp}, the
reconnection probability of the strings is $p\ll 1$ at any
time during the evolution of the network.
However, in our case the small reconnection probability is
due to the fact that the strings in different vacua cannot reconnect.
Therefore, in the macroscopic scale which is much larger than the
distance $d$, the brane 
necklaces look precisely the same as the original $p\ll 1$ strings,
while in the microscopic scale they look quite different.
For example, if the wavelength of the wiggles $l_{wiggles}$ is much
smaller than $d$, the production of the small loops is efficient and 
at least in the local region the brane necklaces look like conventional
$p\sim 1$ strings. 
On the other hand, if the wavelength of the wiggles $l_{wiggles}$ is
much larger than $d$, one can easily understand that the evolution of
the brane necklaces is quite similar to the $p\ll 1$ strings.
Therefore, in the case of $r \ll 1$, we can conclude that the evolution
of the brane necklaces is precisely the same as the conventional $p\ll
1$ strings  as far as 
\begin{equation}
\label{wigled}
l_{wiggles}(t) \gg d(t)
\end{equation}
is satisfied.
The above relation (\ref{wigled}) puts a cut-off time for the original
analysis of the $p\ll1$ strings, which is not important for the GW burst
analysis at the present cosmic time.
The numerical simulations of this kind are difficult, because of 
the two hierarchically different scales.
However, what is important for the observation
of the GW emitted from strings is the behaviour at the macroscopic scale.
Therefore, it must be important to note that at later period the behaviour
of the brane necklaces is quite the same as the conventional $p\ll 1$
strings, which is already discussed in ref.\cite{vilenkin-lowp}.
Thus we can conclude that the brane necklaces can have significant impact on
the observation of the gravitational wave. 
Please be sure that in our model the small reconnection probability is
due to the small angle.

Our next example is the peculiar case of $r \gg 1$.
In this case, the characteristic length scale of the network will be
modified\cite{vilenkin-necklace} as:
\begin{equation}
\xi(t) \sim (r+1)^{-1/2}t.
\end{equation}
The string length per unit volume is $\sim p^{-1}\xi^{-2}$.
The number of loops formed per Hubble volume per Hubble time is
therefore\cite{vilenkin-lowp}
\begin{equation}
N_l \sim \frac{r}{p\alpha}.
\end{equation}
The loop self-intersections should be frequent
and their fragmentation into smaller loops is very efficient, which
suggests that the lifetime of a loop of size $l$ is $\tau_{r\gg1} \sim 
lr^{-1/2}$\cite{vilenkin-necklace}.
In this case, it is easy to see that the loops produced by the necklaces
of $r \gg 1$ are short-lived and the calculation of the loop density is 
similar to the cases of small $\alpha$.
The corresponding loop density is therefore
\begin{equation}
n(t) \sim \frac{r^{1/2}}{ p t^3}.
\end{equation}
Here we have assumed that the wiggles are small compared to the typical
scale of the network, so that $l_{wiggles}<\xi$.
The results obtained here suggests that the analysis of the GW signals
emitted by the network of the cosmic necklaces are similar to the
original argument\cite{vilenkin-lowp}.
The difference that we have found is the suppression factor $\Gamma G
\mu r^{1/2} $ in their loop density.

Therefore, we may conclude that there is no qualitative difference
between the brane necklaces and the conventional strings(or necklaces)
as far as 
the observation of the gravitational wave signals is concerned. 
However, as we have discussed above, chopped loops of the brane necklaces can be
stabilized due to their non-zero windings around compactified
space.
The stable relics of such loops are superheavy, which must satisfy the
cosmological bound of the conventional cold dark matter.
Therefore, in order to find qualitative differences, it is quite
important to calculate the relic density of the brane coils.

To calculate the relic density of the stable relics, it is convenient to
introduce the relative abundance $Y_s =n(t)/s(t)$, where $s(t)$ is the
entropy density.
In our case, small loops are produced through multiple self-interactions
of larger loops, whose sizes are determined by
the wavelength of the wiggles $l_{wiggles} \sim \alpha t$.
The conventional analysis of $p \ll 1$ string network is reliable as far
as $l_{wiggles}\gg d$.
\begin{itemize}
\item
Let us first consider the case where the effective potential $V(\phi)$ is
due to the stringy dynamics.
In this case, monopoles are produced by the phase
transition just after angled inflation.
The number of the loops formed per Hubble volume per Hubble time is
already given in ref.\cite{vilenkin-lowp}.
For $r<1$, it becomes
\begin{equation}
N_l \sim 1/(p\alpha),
\end{equation}
which cooresponds to the number of the coils emitted from the network of the
     brane necklaces.
Therefore, we can write
\begin{equation}
\label{dotn}
\dot{n}_{coil}\sim 1/(p\alpha t^4).
\end{equation}
In the most efficient case when all the loops are stabilized by the
coils, the relic number density of the coils is obtained by 
integrating eq.(\ref{dotn}).
The ratio $Y_s = \int (\dot{n}/s) dt $ is:
\begin{equation}
Y_s \simeq (p\alpha)^{-1}M_p^{-3/2} \epsilon_t^{1/2} t^{-3/2}_{m}
\end{equation}
for $r <1$, and
\begin{equation}
Y_s \simeq (r/p\alpha)M_p^{-3/2} \epsilon_t^{1/2} t^{-3/2}_{m}
\end{equation}
for $r \gg 1$.
Here $t_m$ is the time when the monopoles are produced as the kinks on
     the strings,
and $\epsilon_t$ is defined by $\epsilon_t =t_m /t_{coil}$, where
$t_{coil}$ is the cut-off time when the wavelength of the wiggles becomes
     much larger than the typical distance between monopoles.
We have assumed that the coils produced before
     $t_{coil}$ are  negligible.
\item
We must be careful if the effective potential
$V(\phi)$ is lifted by the conventional mechanism of supersymmetry 
breaking.
In this case, since the sizes of the
monopoles are comparable to the Hubble radius when they are
     formed, initially the wiggles cannot form coils.
The coils are formed at much later period when $l_{wiggles} \gg d$,
which means $\epsilon_t \ll 1$.
\end{itemize}

As we have discussed above, a rough estimation of the characteristic
properties 
of the networks of the brane necklaces is straightforward.
Their evolution can be approximated by the evolution of the conventional
$p\ll 1$ strings or the necklaces.
On the other hand, one may suspect that the ``vacua'' on the strings
might not be degenerated, and the brane necklaces might not behave as we have
discussed above.
We think this claim could be true.
If the vacua on the strings are only quasi-degenerated, the tension of
the strings in the different domains will be different by a small factor
$\Delta\mu$. 
Due to $\Delta\mu \ne 0$, the monopoles may move along the strings,
and may be annihilated or gathered to form coils on the strings.
After the ``true vacuum'' dominates the necklaces, the reconnection
probability of the necklaces will be enhanced up to $p\simeq 1$, because
the other ``false vacua'' have shrunk to a point.
Then the situation becomes quite the same as the evolution of the
conventional $p\sim 1$ necklaces\cite{vilenkin-necklace}.
The only difference is that the monopoles are replaced by the coils, which
can stabilize the loops of the strings.

\section{Conclusions and Discussions}
In this paper we have investigated the cosmological evolution of the cosmic
 strings in angled inflation.
We have shown that the cosmic strings in angled inflation turn out to be
 the brane necklaces. 
The evolution of the networks of the brane necklaces is similar to the
 conventional $p\ll 1$ strings. 
However, the loops of the brane necklaces will shrink to produce superheavy
 winding states,  which look like coils.
In angled inflation, it is therefore possible to distinguish the brane
 necklaces from the conventional cosmic strings and necklaces.   
Moreover, the reconnection probability of the brane necklaces can be
enhanced up to $p \sim 1$ if the degeneracy of the vacua is resolved by
 a small perturbation.
We have made a rough estimation of the characteristic properties of the
 network of the brane necklaces. 
It is obvious that the numerical simulations are required to
 obtain more sensible results.

Here we should discuss what we can say about the conventional brane
strings in the 
simplest scenario 
of $D\overline{D}$ brane inflation.  
As is already discussed in ref.\cite{Polchinski-GW, vilenkin-lowp}, it
seems natural to think that the position of the daughter branes in the
compactified space should be stabilized.
On the other hand, however, there is no sensible reason that we have to 
believe that the potential has only a unique global minimum in the
compactified 
space. 
Therefore, it is always important to consider daughter branes that have
multiple minima in the compactified space.
As a result, it is always natural to consider cosmic strings that have many
domains on their world volume, which induces kinks interpolating between
them. 
Even if the flat directions are stabilized at a unique minimum, winding
states (coils) can appear if the Brownian motion induces
 windings around a nontrivial circle in the compactified space.
Naively, the nontrivial circle means a conventional noncontractible
circle that winds around compactified space. 
In generic cases, however, the nontrivial circle could be a valley of the
potential that winds around a hill.

It is obvious that {\bf any} branes that are produced by
brane collision can have flat directions in the internal space.
Then due to the Brownian motion, the cosmic strings will become the brane
necklaces or the brane coils.
The brane necklaces can therefore be produced in other generic brane models.
If the necklaces are produced, they can be used to probe the internal
structure of the compactified space.
Moreover, future numerical simulations will provide us the important
information about the undetermined parameters, which may put serious
bounds on brane inflation.

\section{Acknowledgment}
We wish to thank K.Shima for encouragement, and our colleagues in
Tokyo University for their kind hospitality.

\end{document}